\documentclass[journal=jacsat,manuscript=article]{achemso}

\usepackage{xcolor}
\usepackage[version=3]{mhchem} 

\DeclareUnicodeCharacter{0301}{\'{e}}

\author{Anastasiia Zalogina}
\affiliation{Nonlinear Physics Centre, Research School of Physics, Australian National University, Canberra ACT 2601, Australia}

\author{Javid Javadzade}
\affiliation{3rd Institute of Physics and Center for Applied Quantum Science, University of Stuttgart, 70569 Stuttgart, Germany}

\author{Roman Savelev}
\affiliation{Department of Physics and Engineering, ITMO University, St. Petersburg 197101, Russia}

\author{Filipp Komissarenko}
\affiliation{Department of Physics and Engineering, ITMO University, St. Petersburg 197101, Russia}

\author{Alexander Uvarov}
\affiliation{Saint Petersburg Academic University, St. Petersburg 194021, Russia}

\author{Ivan Mukhin}
\affiliation{SCAMT Institute ITMO University, St. Petersburg 197101, Russia}
\alsoaffiliation[Second University]
{Saint Petersburg Academic University, St. Petersburg 194021, Russia}
\alsoaffiliation[Second University]
{Higher school of engineering physics, Peter the Great St.Petersburg polytechnic university, St. Petersburg 195251, Russia}

\author{Ilya Shadrivov}
\affiliation{Nonlinear Physics Centre, Research School of Physics, Australian National University, Canberra ACT 2601, Australia}
\email{ilya.shadrivov@anu.edu.au}

\author{Alexey Akimov}
\affiliation{PN Lebedev Institute RAS, 119991, Moscow, Russia}
\alsoaffiliation{Russian Quantum Center, 143025, Skolkovo, Moscow, Russia}
\alsoaffiliation{National University of Science and Technology, Moscow, 119049, Russia}

\author{Dmitry Zuev}
\affiliation{Department of Physics and Engineering, ITMO University, St. Petersburg 197101, Russia}
\email{d.zuev@metalab.ifmo.ru}

\title[An \textsf{achemso} demo]
  {Control of NV center radiation in nanodiamonds by silicon nanoantennas}

\abbreviations{IR,NMR,UV}
\keywords{American Chemical Society, \LaTeX}

\begin{document}








\begin{abstract}
The development of nanophotonics systems for the manipulation of the luminescent properties of single quantum emitters is essential for quantum communication and computing. Dielectric nanosystems enable various opportunities for light control through inherent electric and magnetic resonances, however their full potential has not yet been discovered. Here, the emission properties of NV centers in nanodiamonds placed in the near-field zone of silicon nanoresonators are investigated. It is demonstrated experimentally that the spontaneous emission rate of single NV centers in 50~nm nanodiamonds can be modified by their coupling to spherical nanoantennas, reducing the mode of the lifetime distribution by $\approx2$ times from 16~ns to 9~ns. It is also shown that the collected intensity of photoluminescence emission from the multiple NV centers in 150~nm nanodiamond coupled to a cylindrical nanoantenna is increased by more than 50\% compared to the intensity from the same nanodiamond on a bare substrate.

\end{abstract}

\section{Introduction}

Tailoring the luminescent properties of single quantum emitters by their coupling to specially-designed photonic nanostructures is a cornerstone of the development of efficient single photon sources for quantum communications and quantum computing. Among various platforms, solid-state single-photon sources have attracted considerable attention in nanophotonics technologies and have been the focus of extensive research in the past decades~\cite{norman2020novel, zhang2020material, atature2018material}. Color centers in diamonds are of particular interest due to their unique properties: they can be implemented in quantum optical devices functioning even at room temperature, they exhibit highly coherent optical transitions~\cite{balasubramanian2019dc, wan2020large} and provide the possibility of the optical initialization, manipulation and readout of spin states~\cite{wehner2018quantum}. 

Among various color centers in diamond, NV centers are the most robust and well-studied systems that can operate as single-photon sources at room temperature~\cite{aharonovich2016solid, bradac2019quantum}. The key challenge that prevents the practical implementation of these defects in practical applications is their relatively low brightness. Conventionally, this issue is addressed by coupling the nanodiamonds with color centers to optical nanostructures, which can increase their emission and excitation rates and make emission more directional or channel it into a guided structure~\cite{siampour2018chip, siampour2017nanofabrication, geiselmann2014deterministic, schietinger2009plasmon}.

One of the advantages of using color centers in diamonds is that they can be realized in the form of a small (of the order of ten nanometers) diamond nanoparticle, which is suitable for precise positioning in a specific place near the designed optical nanostructure. This was already demonstrated for various photonic structures, such as photonic crystals~\cite{fehler2019efficient}, solid-immersion lenses~\cite{huang2019monolithic}, plasmonic~\cite{parzefall2017antenna, szenes2017improved} and dielectric~\cite{staude2019all, makarov2019halide, sanz2018enhancing, lukin20204h} nanoantennas. High-index dielectric nanoparticles are of particular interest, as they demonstrate both electric and magnetic dipole responses and have small material losses in the optical spectral range~\cite{staude2019all}. For example, it was recently demonstrated that diamond resonant nanoparticles are able to modify emission properties of the embedded NV centers~\cite{zalogina2018purcell, shilkin2017optical, obydennov2021spontaneous}. Further development of this approach towards its implementation into realistic practical devices requires engineering and research into systems which combine diamond nanoparticles with color centers and resonant structures made of high-index semiconductor materials.     

\begin{figure*}
 \centering
 \includegraphics[height=6cm]{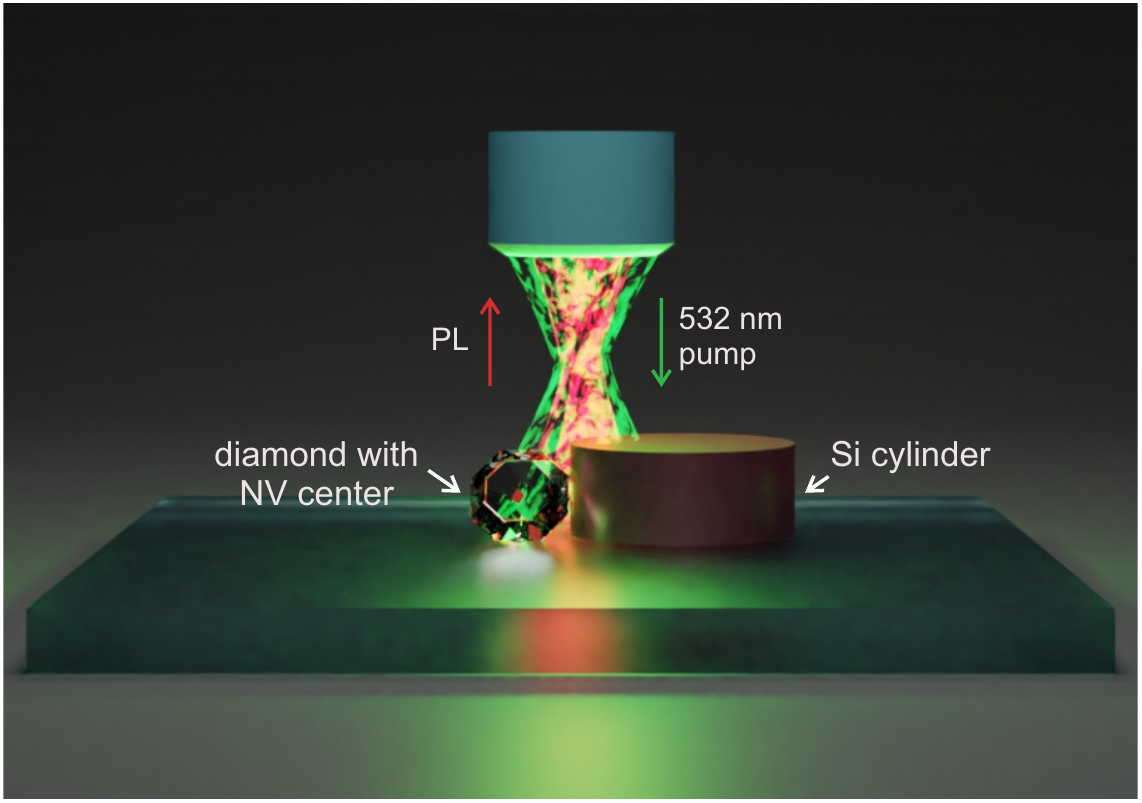}
 \caption{{\bf Schematic representation of a nanodiamond with embedded NV center coupled to a cylindrical silicon nanoantenna.}}
 \label{Figure_1}
\end{figure*}

In this work, the emission properties of NV centers embedded in nanodiamonds coupled to resonant silicon nanoantennas are experimentally studied, as illustrated schematically in Figure~\ref{Figure_1}. Two alternative approaches for the fabrication of these ``nanodiamond-nanoantenna'' structures are employed. The first one relies on the creation of a random distribution of nanodiamonds with linear size $\approx$50~nm containing single color centers placed near spherical nanoantennas. Measurements performed for these samples have shown that $the~variation$ in the NV centers' lifetime becomes approximately two times larger than the lifetime of the NV centers in nanodiamonds on a glass substrate covered with ITO, while at the same time the mode of the lifetime distribution reduces by $\approx2$ times from 16~ns to 9~ns. The second approach relies on the precise positioning of a nanodiamond containing multiple NV centers close to a nanoantenna. In this approach a nanodiamond with a larger size of $\approx$150~nm is chosen which was precisely positioned and coupled to a cylindrical silicon nanoantenna {\em via} the pick-and-place method. This made possible to compare the emission properties for the same nanodiamond with and without a nanoantenna. The measurements have shown that the effect of the nanoantenna on the lifetime of the NV centers is negligible, while due to angular redistribution of the radiation, the fraction of emission power into upper half space is increased by $\approx$50\%.

\section{Fabrication approaches}

\begin{figure*}
 \centering
 \includegraphics[height=9cm]{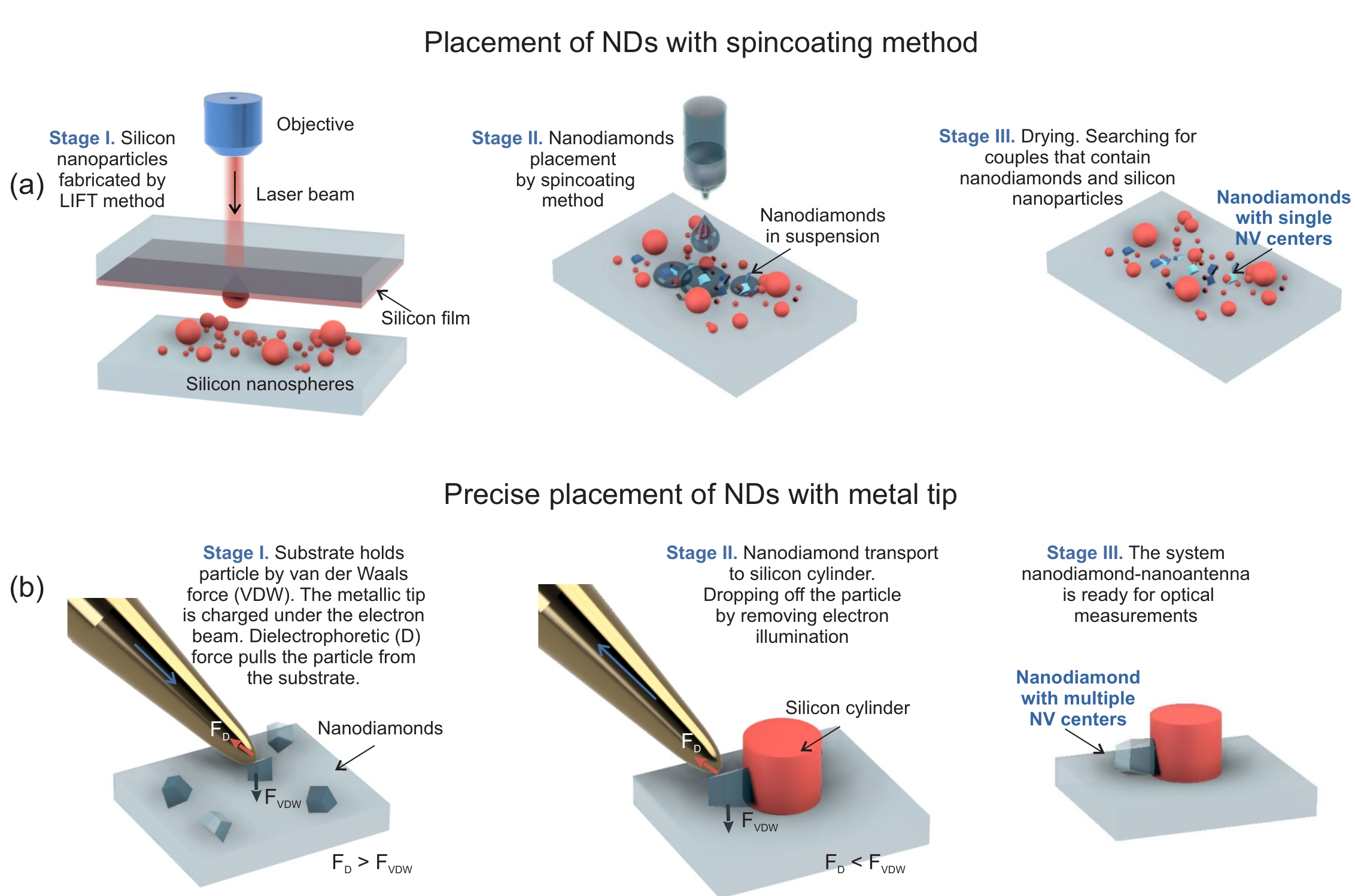}
 \caption{{\bf Schematic illustration of the experimental procedures. (a) The structure consists of a nanodiamond containing single NV center coupled to a spherical silicon nanoparticle. (b) The structure consists of a nanodiamond containing multiple NV centers coupled to a cylindrical silicon nanoparticle.}}
 \label{Figure_2}
\end{figure*}

The two approaches employed in the study for the fabrication of silicon nanoantenna-diamond nanoparticle systems are schematically illustrated in Figure~\ref{Figure_2}. The first approach is based on spin-coating deposition of nanodiamonds with single NV centers on top of spherical silicon nanoantennas and it consists of three main steps. The first step (Stage I in Figure~\ref{Figure_2}a) is the production of silicon nanoantennas of approximately spherical shape and random size distribution by conventional laser printing using a femtosecond laser at a wavelength of 1050~nm irradiating thin Si film on glass. For the second step (Stage II in Figure~\ref{Figure_2}a), a suspension containing nanodiamonds specified as 50~$\pm$~10~nm by the manufacture with embedded NV centers was used. The nanodiamonds were deposited with the spin-coating method. Finally, after drying the sample under a nitrogen flux, nanodiamond/nanoantenna pairs that exhibited single photon emission were searched with scanning electron microscope (SEM) (Stage III in Figure~\ref{Figure_2}a).  
Due to the relatively small sizes of the nanodiamonds many of them contained single NV centers, and at the same time they were easily dispersed onto the substrate. Although such an approach does not allow for a comparison of the emission properties of the same emitter with and without a nanoantenna, it produces a vast amount of statistical data from the measurements of nanodiamond-nanoantenna systems that allows the identification of the average effect produced by the silicon nanoantennas on the emission properties of the single NV centers.

The second approach utilizes the so-called ``pick-and-place'' method for precise positioning of a specifically chosen diamond nanoparticle near a specifically chosen nanoantenna~\cite{denisyuk2014electrostatic, denisyuk2014mechanical}. Although such an approach allows the control of both nanoantenna and nanodiamond properties before combining them into a single active nanostructure, the complexity of this approach limits the number of realizations and makes the amount of experimental data much smaller. In the first step of this approach, amorphous silicon cylindrical nanoantennas with given parameters were manufactured using of electron beam lithography and plasma etching of a thin layer of amorphous silicon. The nanoantennas were further annealed with a femtosecond laser to achieve the crystalline phase of the silicon (see Supporting Information for details). In the second step, a 150~$\pm$~25~nm (estimated from SEM) nanodiamond containing multiple NV centers was picked up from the glass substrate with a sharp tungsten tip inside the SEM (Stage I in Figure~\ref{Figure_2}b) and transferred to another substrate with the fabricated array of silicon cylinders (Stage II in Figure~\ref{Figure_2}b). Finally, the nanodiamond was detached from the tip by reducing the electric charge of the tip.

\subsection{Theoretical background}
Placement of a nanodiamond with embedded color centers in the vicinity of a nanoantenna supporting Mie resonances can increase the number of photons emitted by the color centers and collected in the far zone. Such a modification can be related to an increase in the decay and excitation rates, and improvements in the directionality of the emission. The first factor -- change of the total decay rate, i.e. the Purcell factor~\cite{purcell1995spontaneous} -- can be measured by the conventional time-correlated single-photon counting method (TCSPC). Since NV centers have a broadband emission spectrum, experimental measurements allow us to observe only the spectrally-averaged Purcell effect, because an increase in the local density of optical states in a narrow frequency range due to the presence of a resonance in a silicon nanoantenna will affect only that particular part of the whole emission spectrum. Consequently, the dimensions of the silicon nanoantennas have a minor influence on the decay rate as long as the studied particles are large enough to support a few resonances within the NV centers' luminescence spectral range. The same reasoning can be applied to the excitation rate due the bandwidth of the absorption in NV centers being approximately the same.

The change in the number of photons collected from the NV centers in the objective of a given aperture due to their coupling to a nanoantenna can be even more complex. It is determined by the sum of the radiation patterns of all modes excited in the nanoantenna with weights that depend on both the near-field distribution of the modes (which defines the coupling strength) and the luminescence spectrum of NV centers in bulk diamond, and the far-field pattern of each particular mode. Considering that the emitter orientation and position strongly influence the coupling strength~\cite{lai2010optical, dolan2014complete, epstein2005anisotropic}, accurate theoretical calculation of such an effect for a particular sample becomes practically impossible due to many the unknown experimental parameters. Therefore, it is difficult to achieve a complete theoretical understanding of these systems. However, useful results can be derived from statistical data gathered from experiments.   

\section{Results and Discussion}

\begin{figure*}
 \centering
 \includegraphics[height=10cm]{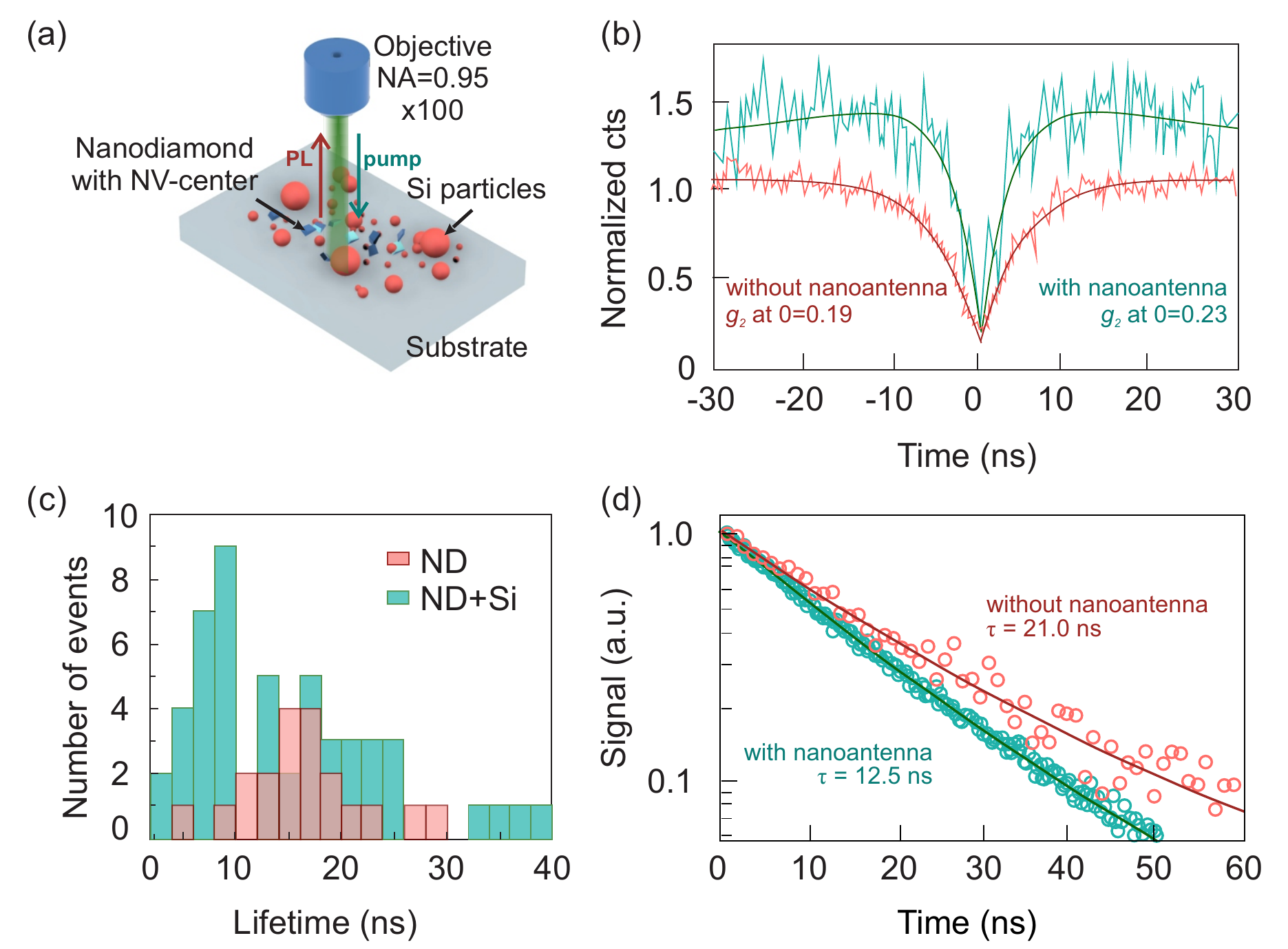}
 \caption{{\bf 
 (a) Schematic illustration of the experimental method used to search for nanodiamonds containing single NV centers located in the vicinity of the silicon nanoparticles, and to measure their optical properties. (b) Second-order correlation function $g_{2}(t)$ of the luminescence signal from the NV centers in a nanodiamond with (green curve) and without (red curve) a silicon nanosphere. (c) Histogram of the lifetimes of the NV centers in nanodiamonds on a bare substrate (red color) and coupled to silicon nanoparticles (green color).
 (d) Normalized photoluminescence decay curves for nanodiamonds with (green curve) and without (red curve) a silicon nanosphere. }}
 \label{Figure_4}
\end{figure*}
 
In the first of the two experimental approaches (Figure \ref{Figure_2}a), a sample containing randomly-distributed spherical silicon nanoantennas with radii in the range 140-160~nm mixed with diamond nanoparticles of mean radius 50~nm is fabricated. A schematic demonstration of the subsequent measurement experiment is shown in Figure~\ref{Figure_4}a. In order to measure the effect of nanoantenna on the emission properties of the NV centers, an additional substrate containing spin-coated diamond nanoparticles on a bare substrate without silicon nanoantennas is fabricated. The number of NV centers in the diamond nanoparticles was random, therefore the diamond nanoparticles that exhibit single-photon emission is firstly identified. This was done by measuring the second-order correlation function $g^{(2)}(\tau)$ using the Hanbury Brown and Twiss configuration~\cite{hanbury1956question}. An example of the measured $g^{(2)}(\tau)$ with a pronounced antibunching dip at $\tau=0$ is shown in Figure~\ref{Figure_4}b for a nanodiamond near a nanoantenna and for a nanodiamond on a bare substrate. From the available samples, those with measured values of $g^{(2)}(0)$ below 0.5 were selected, which indicated their single-emission character. It is worth noting that the presence of a nanoantenna increases the pumping efficiency, which affects the profile of the $g^{(2)}(\tau)$ curve as shown in Figure~\ref{Figure_4}b. Overall 50 nanodiamonds with single NV centers coupled with silicon nanoantennas, and 20 nanodiamonds on the bare substrate are identified.

Next lifetime measurements for the chosen nanodiamonds are performed using the TCSPC method. The obtained results are presented in Figure~\ref{Figure_4}c.
The lowest measured lifetime of an NV center near a silicon nanosphere was $\approx$3~ns, while the peak value reduced to $\approx9$~ns. This may be compared to the nanodiamonds on the bare substrate, for which the measured lifetime was in the range $5-30$~ns with a peak value of $\approx16$~ns. This is a reduction of the mode of the lifetime distribution of $\approx2$ times, which demonstrates that the spontaneous emission rate of a single NV centre can be modified by coupling it to a spherical silicon nanoantenna. The spread of the lifetime values is most likely caused by different dipole orientations and positions of the NV centers inside the nanodiamonds i.e.\ their proximity to the nanoantenna as well as to the substrate. Examples of the photoluminescence decay curves for 50~nm nanodiamond with single NV centers near spherical silicon nanoantennas are shown in Figure~\ref{Figure_4}d. This demonstrates a significant decrease in the lifetime from 21~$\pm$~2.1~ns to 12.5~$\pm$~1.3~ns, due to the presence of the nanoantenna.

\begin{figure*}
 \centering
 \includegraphics[height=7.3cm]{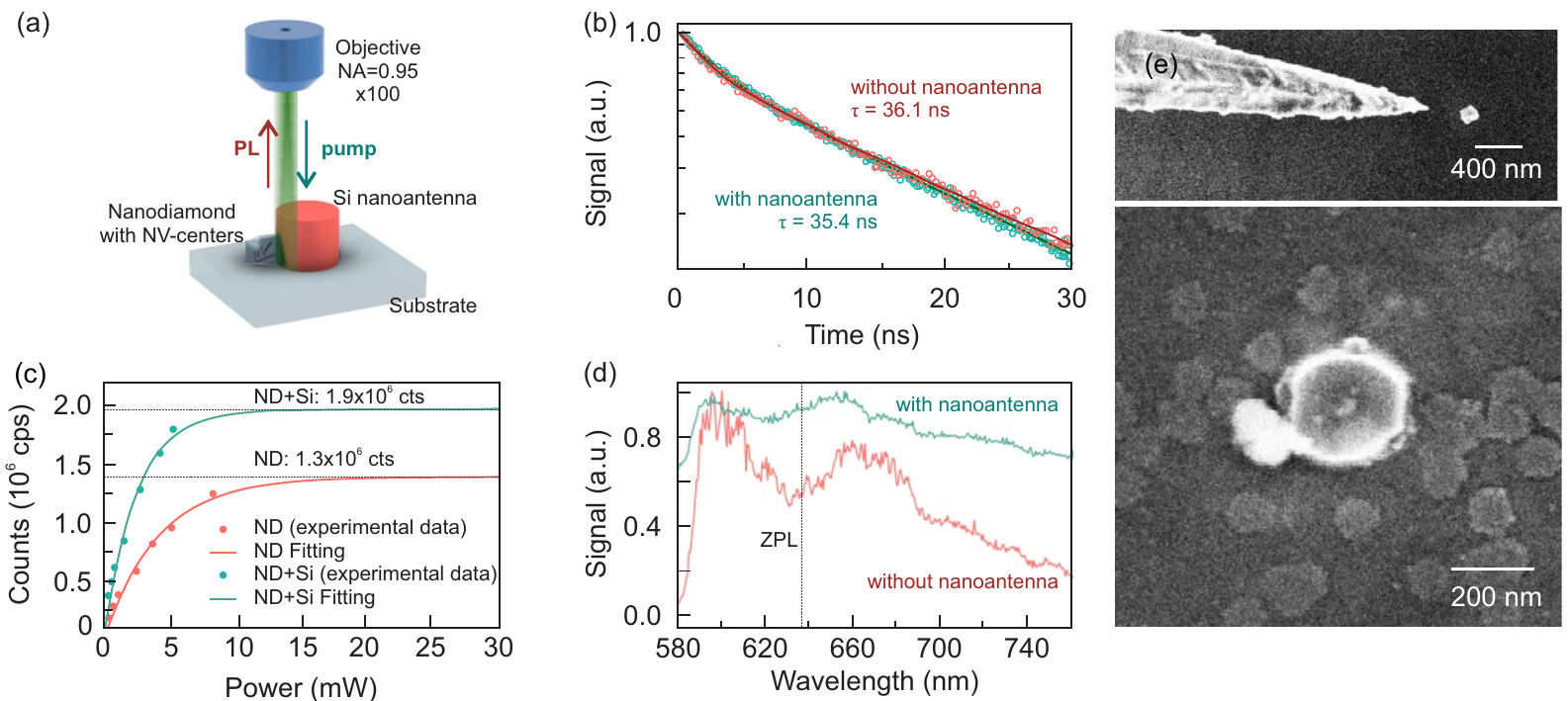}
 \caption{{\bf
(a) Schematic illustration of the system under study: a nanodiamond containing multiple NV centers placed in the vicinity of a cylindrical silicon nanoantenna. (b) Normalized photoluminescence decay curves, (d) saturation measurements,  and (e) dark-field spectra of a nanodiamond without (red curves) and with nanoantenna (green curves); in (d) points correspond to the experimental data, and solid curves to the fitting functions. ND denotes 'nanodiamond'. (e) SEM images demonstrating the "pick and place" method: a nanodiamond is picked up (above), transferred to another substrate and placed near the silicon cylinder (below).}}
 \label{Figure_5}
\end{figure*}

In the second approach (Figure \ref{Figure_2}b), cylindrical silicon nanoantennas with diameters of $\approx$270~nm and heights of $\approx$220~nm is firstly fabricated as described in the Methods section. The transfer of a nanodiamond from a glass substrate to the nanoantenna was realized using a sharp tungsten tip inside the SEM. Since such a process is highly delicate and almost impossible if the size of the moved object is $\leq$ 50~nm, a nanodiamond with $\approx$150~nm size containing multiple NV centers is chosen. Figure~\ref{Figure_5}e shows micrographs of this process: first the collection of the nanodiamond from the substrate, then its deposition near the $\approx150$~nm nanoantenna.

Next, the measured luminescent properties of the chosen nanodiamond on a bare substrate and near the nanoantenna are compared. Three characteristics are measured: the lifetime of the NV centers measured using the TCSPC method (Figure~\ref{Figure_5}b), the dependence of the total number of emitted photons on the pump power (Figure~\ref{Figure_5}c), and dark field spectra collected with an objective with NA=0.95 and angle of collection 72.1 degrees (Figure~\ref{Figure_5}d).

The measured lifetime of the NV centers shown in Figure~\ref{Figure_5}b demonstrates that the nanoantenna has a minor effect on the emission rate. This relatively small effect on the emission rate, compared to the results using spherical nanoantennas, is likely due to the much larger size of the nanodiamond which reduces the Purcell factor. The randomly-distributed nature of the NV centers may also reduce the Purcell factor. In the work \cite{zalogina2021comparison} it was theoretically shown that an increase of nanodiamond sizes from 30~nm to 150~nm led to a fourfold decrease in the Purcell factor. 

In order to compare the emitted power with and without a nanoantenna, the experimental data is fitted (Figure~\ref{Figure_5}d) with the ansatz $I_{}=\frac{I_{sat}\cdot P}{P_{sat} + P}~,$ where $I$ is the measured intensity, $I_{sat}$ is the saturation intensity, $P_{sat}$ is the saturation pump power, and $P$ is the pump power. The results reveal that there is a noticeable change in the signal intensity from the NV centers in the presence of the nanoatenna. The number of counts per second in the saturation regime for the nanodiamond on a bare substrate was 1.3x10$^{6}$, while near the nanoantenna the number of counts per second increased to 1.97x10$^{6}$. Such an increase could be attributed to a change in the emission rate and/or a change in the radiation pattern. This increase demonstrates that the fraction of the power collected in the experiment was increased by $\approx$1.5 times by coupling of the NV centers to the resonances of the nanoantenna. This coupling is accompanied by a noticeable reshaping of the dark-field spectrum (Figure~\ref{Figure_5}e).

\section{Conclusion}

In this work, the experimental study of emission properties of the NV centers in diamond nanoparticles coupled to the silicon nanoantennas have been presented. In the first part it was shown that laser-ablated crystalline silicon nanospheres supporting resonances are able to modify the emission rate of single NV centers inside 50~nm nanodiamonds. It is demonstrated that the the peak lifetime value of single NV centres inside diamond nanoparticles in the vicinity of spherical silicon nanoparticles is reduced to $\approx9$~ns, as compared to nanodiamonds on a bare substrate, for which the measured peak value is $\approx16$~ns. An increase in the spread of the spontaneous emission rate distribution, and a shift of its peak value, were observed and are attributed to a modification of the local density of optical states in the proximity of the nanoantenna.  

In the second part, cylindrical silicon nanoantennas were fabricated using electron beam lithography and a 150~nm nanodiamond with multiple NV centers was placed precisely near a nanoantenna using a sharp tungsten tip inside the SEM. In this case silicon nanoantennas were able to produce an increase in the number of counts from the nanodiamond by 1.5 times compared to without the nanoantenna. This effect can be attributed to a change in the directivity of the NV centers' emission. 

These results are motivation for further studies of the spontaneous emission rate enhancement of color centers in nanodiamonds by their coupling to nanoantennas made of high-index materials. We believe that more pronounced emission rate enhancement and radiation pattern modification can be achieved with more accurate control of the NV centers and their precise placement inside a nanodiamond. 

\subsection{Methods}

\subsection{Spherical Silicon Nanoparticles Fabrication}

Silicon nanoparticles were fabricated using the method of laser ablation (Stage I in Figure~\ref{Figure_2}a). To get the nanoparticles with this method, the pulsed Yb$^{3+}$ femtosecond laser with the wavelength of 1050~nm, the repetition rate of 80 MHz and the energy $<2$ nJ (TEMA-100, Avesta Project). The pulses are selected with the help of Pockels cell-based pulse picker (Avesta Project).   The laser beam is focused with the objective magnification of 10X and numerical aperture NA=0.26. For the forward transfer of nanoparticles an 80-nm thin film of amorphous silicon ($a-Si:H$) is proposed to use. The silicon film is deposited on a glass substrate by plasma enhanced chemical vapor deposition from the precursor gas $SiH_{4}$.  The receiving substrate is located under the silicon film with several micrometers space between them. Since the amorphous silicon possesses the imaginary part of refractive index, which is almost twice larger in the visible spectrum range compare with crystalline silicon~\cite{palik1998handbook}, it is essential to check the structure of silicon nanoparticles.  The crystalline structure of silicon nanoparticles is examined using Raman spectroscopy. The printed silicon nanospheres are characterized with a crystalline cubic structure with the narrow peak at 521.5~cm$^{-1}$, which are observed at the Raman spectra. For studying the size and morphology of silicon nanoparticles the scanning electron microscope (SEM) is used (Neon 40 EsB, Zeiss). 

\subsection{Cylindrical Silicon Nanoparticles Fabrication}

Cylindrical silicon nanoantennas were fabricated with electron beam lithography. In the first step, a 220~nm film of hydrogenated amorphous silicon ($\alpha$-Si:H) was deposited onto a quartz substrate using plasma-enhanced chemical vapor deposition (PECVD). In the next step, the pattern was defined {\em via} electron beam lithography and following by the deposition of metal. The resist layer was lifted off and the pattern was transferred onto the material using an inductively-coupled plasma technique in the presence of SF$_{6}$ and O$_{2}$ gases and at a temperature of 256~K. 
The formation of nanostructures is completed by the removal of the metal layer in a solution of KI:I$_{2}$. The result is silicon nanoantennas in the form of cylinders.

\subsection{Cylindrical Silicon Nanoparticles Crystallization}
Crystallization of cylindrical silicon nanoantennas was realized in order to increase the refractive index of material. An array of nanoparticles was annealed by femtosecond laser (Femtosecond Oscillator TiF–100F, Avesta Poject) at the wavelength of 1045~nm ($\pm$15nm) and power of 200-440~mW, and using objectives Mitutoyo M Plan Apo IR for the excitation - 10x/0.28, and 50x/0.45 for collection. During annealing the Raman signal was collected as well, where we were able to see the modification of spectra in real time. The broad peak of amorphous silicon spectra at 477~$cm^{-1}$ was narrowed and shifted to 520~$cm^{-1}$, which indicated crystalline phase of silicon (see Figure S1 in supplementary).  

\subsection{Nanodiamonds with NV Centers}

Nanodiamonds with single NV centers were obtained from Microdiamond AG. In the first approach with spherical silicon nanoantennas the nanodiamonds with linear size 50~$\pm$~10~nm (according to manufacturer) were used. In the second approach with cylindrical nanoantennas diamond nanoparticle with size 150~nm~$\pm$~25~nm (estimated from SEM image) was used. Suspensions that contain nanodiamonds and distilled water were placed onto substrate and spincoated at 2000 rpm for 2 minutes.

\subsection{Lifetime Measurements}
The measurements of lifetime are carried out by the time-correlated single-photon counting (TCSPC) method. TCSPC method determines the lifetime measuring the time between the NV centers excitation and the emitted photons arriving to the detector~\cite{lakowicz2013principles}. NV centers in nanodiamonds are excited with 532-nm pulsed laser (PicoQuant;LDH-P-FA-530XL). The sample is scanning by a laser beam using galvano mirrors (Cambridge Technology; 6215H). Photoluminescence signal is collected with objective Olympus (100$\times$, NA$=$0.9). The collected signal is separated from the excitation beam by the dichroic mirror (Semrock; LPD01-633RU-25), notch filter (Semrock; NF03-532E-25) and longpass filter (Semrock; LP02-633RU-25).  For the detection of NV centers emission, a single-photon avalanche photodiode (SPAD) (PerkinElmer; SPCM-AQRH-14-FC) is used. The SPAD signal is processed by computing counting board (National Instruments; BNC-2121) and the electronic correlation module (PicoQuant; PicoHarp 300). The experimentally obtained curves are fitted by a tri-exponential function:
\begin{equation}
I(t)~=~A_{0}+A_{1}\exp(-t/\tau_{1})+A_{2}\exp(-t/\tau_{2})+A_{3}\exp(-t/\tau_{3}), 
\end{equation}
where $A_{1}$, $A_{2}$, $A_{3}$, $\tau_{1}$, $\tau_{2}$, $\tau_{3}$ and $A_{0}$ are the fitting amplitudes, decay times and background intensity, respectively.
The photoluminescence lifetime is calculated with the following formula:%
\begin{equation}
\left\langle \tau \right\rangle~={\sum\limits_i A_{i}\tau_{i}^2}/{\sum I_{i}\tau_{i}} 
\end{equation}

There are 20 lifetime measurements of nanodiamonds without silicon nanoparticle and 50 lifetime measurements with silicon nanoparticles. 

\subsection{Antibunching Measurements}

 For the antibunching measurements, the Hanbury Brown and Twiss spectrometer configuration is used. This set-up is based on the same confocal microscope as for the TCSPC method, but with extra SPAD, 50/50 beamsplitter and continuous laser source (Coherent; Compass 315M-100) substituting pulsed one. ($g_{2}(t)$) is measured obtaining the background signal. This signal is corrected with coincidence count rate and  normalized by the total counts on each detector, time-bin width and data collection time. The experimentally obtained curves with antibunching dip are fitted by a biexponential function $\exp(-[\Gamma+R]|t|)$, where $\Gamma$ is the total decay rate and $R$ is is the pump rate.

\subsection{Optical Characterization}
The photoluminescence spectra of NV centers and Raman spectra of silicon nanoparticle are obtained at room temperature on a confocal spectrometer Horiba LabRam HR with a cooled CCD camera (Andor DU 420A-OE 325). 600~g~mm$^\textrm{-1}$ and 150~g~mm$^\textrm{-1}$ diffraction gratings were applied for Raman and PL/dark-field spectra registration, respectively.
For photoluminescence measurements, we excite the samples by the supercontinuum source Fianium SC400-6 with an integrated tunable filter yielding laser pulses with the wavelength of 530~nm, repetition rate of 60~MHz, a pulse duration of 6~ps, and the bandwidth of 20~nm. The He-Ne laser (632.8~nm) and the objective Mitutoyo Plan Apo VIS (100$\times$, NA$=$0.9) are utilized for the Raman spectroscopy. 
The objective Mitutoyo Plan Apo VIS (10$\times$, NA$=$0.28) for the photoluminescence excitation and Mitutoyo Plan Apo NIR (50$\times$, NA$=$0.42) are used for the photoluminescence excitation and dark-field signals collection, respectively. The objective Mitutoyo Plan Apo NIR (10$\times$,NA$=$0.26) is used in a dark-field scheme for a sample illumination at the angle of 67$^{\circ}$ to the surface normal by a white light source (HL-2000 halogen lamp). The  micrometer translation stage (Thorlabs; MBT616D) and atomic force microscope stage (SmartSPM AIST-NT) are used for the sample positioning, which was monitored by the CCD camera (Cannon 400~D).

\begin{acknowledgement}

The authors thank Vadim Vorobyov for discussions. The pick and place experiments, fabrication of nanoantennas by nanolithography as well as SEM studies were supported by the Russian Science Foundation (Grant MD-18-37-00384). The laser-assisted nanofabrication as well as optical experiments were supported by Russian Science Foundation (Grant 21-72-30018).

\end{acknowledgement}

\begin{suppinfo}

\section{Cylindrical Silicon Nanoparticles Crystallization}
Crystallization of cylindrical silicon nanoantennas was realized in order to increase the refractive index of material. An array of nanoparticles was annealed by femtosecond laser (Femtosecond Oscillator TiF–100F, Avesta Poject) at the wavelength of 1045 nm ($\pm$15nm) and power of 200-440 mW, and usuing objectives Mitutoyo M Plan Apo IR for the excitation - 10x/0.28, and 50x/0.45 for collection. During annealing the Raman signal was collected as well, where we were able to see the modification of spectra in real time. The broad peak of amorphous silicon spectra at 477 $cm^{-1}$ was narrowed and shifted to 520 $cm^{-1}$, which indicated crystalline phase of silicon (see Figure 1 in supplementary).

\begin{figure*}
 \centering
 \includegraphics[height=6cm]{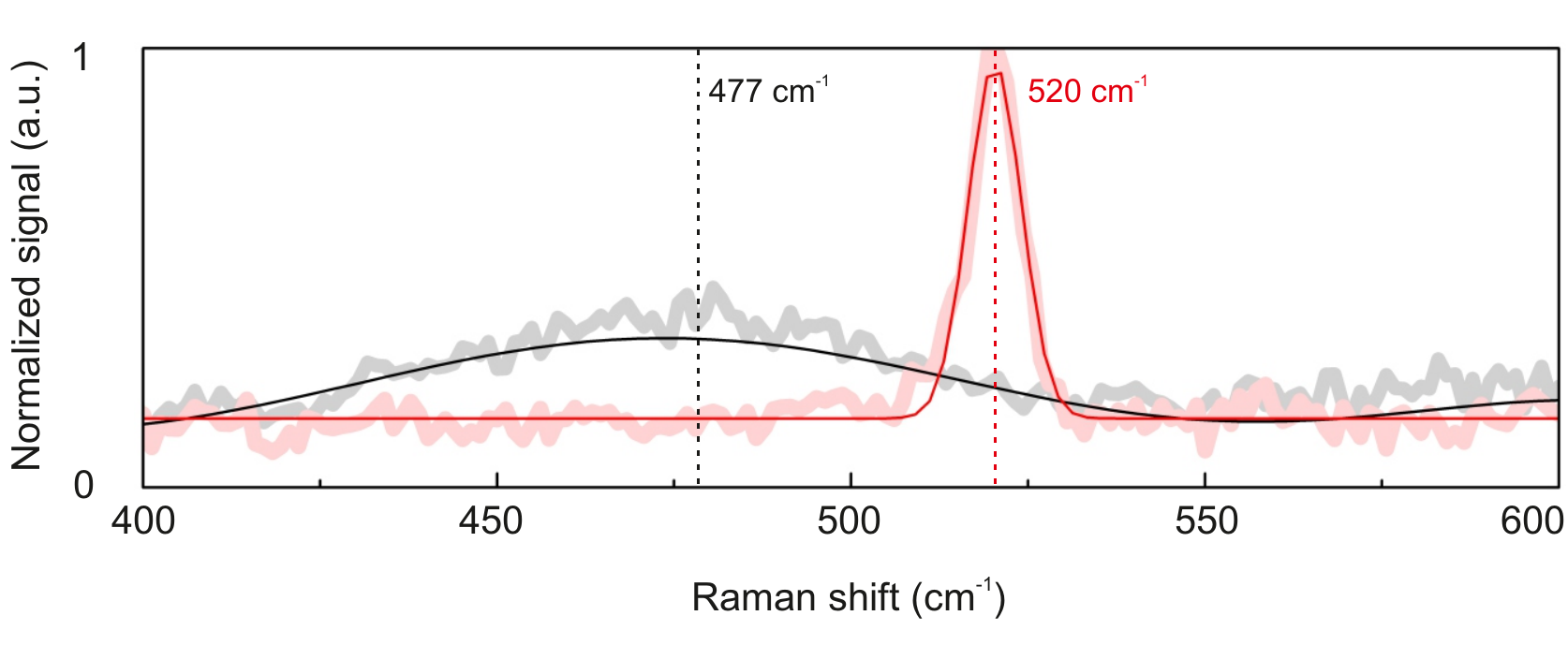}
 \caption{{\bf Raman scattering of Si cylinder before annealing - broad peak for amorphous silicon cylinders (black line) and after annealing - the narrow peak at 520 $cm^{-1}$ of crystalline silicon phase (red line). }}
 \label{Sup_Figure_1}
\end{figure*}

\end{suppinfo}

\bibliography{achemso-demo}

\end{document}